\documentclass[preprint]{revtex4}
\usepackage{graphicx}
\usepackage{dcolumn}
\usepackage{bm}
\begin{document}

\title{Bound and Trapped Classical States of an Electric Dipole in Magnetic Field}
\author{Paulina I. Troncoso and Sergio Curilef}

\address{Departamento de F\'\i sica, Universidad Cat\'olica del Norte, \\ Av. Angamos 0610, Antofagasta, Chile.\\
scurilef@ucn.cl}
\date{\today}
\begin{abstract}
In the present work, we study the classical behavior of an electric
dipole in presence of an external uniform magnetic field. We derive
equations and constants of motion from the Lagrangian formulation.
We obtain an infinitely periodic effective potential that describes
a rotational motion. The problem is not directly separable in
relative and center of mass variables; even though, we are able to
write the energy of the system as a function of an only term, the
relative variable. We define another constant of motion, which
couples the relative with the center of mass variables. We describe
conditions for bound states of the dipole. In addition, we discuss
the problem in the approximation of small oscillations. Finally, we
explore the existence of a possible family of trapped states in a
region of the space where there are no classical turning points.
\end{abstract}

\maketitle \section{Introduction} Nowadays, fundamental elements in
scale of molecular
machinery\cite{science306,science281,nat406_605,nat406_608,nat401_150,nat401_152}
take the attention from several specialists, due to the development
of nanotechnology. An alternating electric field has already used to
explore electronic structures, however this process can be useful to
exchange the orientation of molecules; therefore, it is possible to
obtain some controlled molecular motion by oscillating electric
fields\cite{nat406_608}. Devices on the molecular level are obtained
from the conversion of energy into controlled motion; regardless of
this, it is difficult to repeat this process in a mechanical
molecular motor. This process is common in biological
systems\cite{nat406_605}. By moment, it is expected to find physical
principles of a motor in molecular scale through the dynamics of
rotors in two dimensions. Such rotors are modeled as electric
dipoles in electric or magnetic fields.

It is well known, we know that the simplest many-body system is the
two-body system and we expect to separate the two-body problem in
two problems of a single body. The problem of two equal charges in
an external magnetic field fulfills to this requirement completely.
In a suitable formulation, two independent variables are considered
to describe the motion of the center of mass and the relative
motion, respectively. The classical\cite{ajp65} and
quantum\cite{pla269} physics of the problem are clearly described in
details . On other hand, the problem of two particles with opposite
charges into a magnetic field does not allow to make the separation
of variables in independent equations, as it is possible to do it in
the case presented above. Nevertheless, two constants of motion are
obtained from a suitable Lagrangian formulation. It is possible to
write an equation for the energy in terms of the relative variable
and other constant of motion connecting the relative variable with
the center of the mass coordinate. Classical\cite{ajp65} and
quantum\cite{jpa2} results were previously reported.

In addition, other related cases are discussed in precedent works
\begin{itemize}
\item the dynamics of charges in oscillating electric and magnetic
fields in Ref.\cite{BJP29,pra59},
\item  the dynamics of a dipole in a magnetic field on $Y$ direction,
the motion of its center of mass is restricted to the $Z$ direction,
and its rotation to the $XY$ plane, is discussed in
Ref.\cite{Pursey}.
\end{itemize}
The main goal in this work is to describe the classical behavior of
an electric dipole in an external magnetic field. With this purpose,
we introduce a model for an electric dipole as two particles of
arbitrary masses, where separation between them is constant, and its
charges are equal and opposite. That case is a two-body problem and
its exact analysis requires a difficult treatment and the further
useful understanding, which is obtained by approximations.

The paper is organized as follows. Firstly, we present a general
formulation for an electric dipole in magnetic field. We define the
Lagrangian function and we derive constants and equations of motion.
Secondly, we present the solution for a particular problem; this is,
the planar motion of an electric dipole in a perpendicular magnetic
field. We discuss bound and trapped states. Finally, we summarize
the formulation and main results.

\section{Lagrangian Formulation: Symmetric Gauge}
In the present model, internal coupling holds the two charges of the
dipole together and the Coulomb interaction between the charges
comes to be constant. Then, we consider a rigid dipole, two fixed
charges by a massless rod, in the presence of a uniform magnetic
field. One of particles carries charge $+e$, while the other $-e$.
The magnetic field is derived from a vector potential $\vec{A}$, as
follows $\vec{B}=\nabla \times \vec{A}$. We relate to the particle
$1(2)$ the position ${\vec{r}}_1({\vec{r}}_2)$, the velocity
$\dot{\vec{r}}_1(\dot{\vec{r}}_2)$ and the mass $m_1(m_2)$. The
Lagrangian formulation leads to the following expression:
\begin{equation}
L(\vec{r}_1,\vec{r}_2;\dot{\vec{r}}_1,\dot{\vec{r}}_2)=\frac{1}{2}m_1{\dot{\vec{r}}_1}^2
+\frac{1}{2}m_2{\dot{\vec{r}}_2}^2-\frac{e}{c}\vec{A}(\vec{r}_1)\cdot\dot{\vec{r}}_1
+\frac{e}{c}\vec{A}(\vec{r}_2)\cdot\dot{\vec{r}}_2+\frac{e^2}{\kappa|\vec{r}_2-\vec{r}_1|}
\label{L1}
\end{equation}
where $\kappa $ is the dielectric constant of the medium in which
the motion of the dipole occurs. We define the vector potential
$\vec{A}$ through the symmetric gauge as follows
\begin{equation}\vec{A}(\vec{r}_i)=\frac{1}{2}\vec{B}\times\vec{r}_i,\mbox{       for $i$=1, 2},\label{gaugei}\end{equation}
where $\vec{B}$ is the uniform magnetic field. Now, we take into
account the following change of variables
\begin{equation}\vec{r}=\vec{r}_2-\vec{r}_1,\label{rel} \end{equation} \begin{equation} \vec{R}=\frac{m_1\vec{r}_1 +
m_2\vec{r}_2}{m_1+m_2},\label{CMR}\end{equation} where $\vec{r}$ is
the relative position and $\vec{R}$ is the position of the center of
mass. Now, if we replace Eq.(\ref{gaugei})-Eq.(\ref{CMR})into
Eq.(\ref{L1}), we obtain the following function
\begin{eqnarray}
L(\vec{R},\vec{r};\dot{\vec{R}},\dot{\vec{r}})&=&
\frac{1}{2}M{\dot{\vec{R}}}^2 +\frac{1}{2}\mu{\dot{\vec{r}}}^2 +\frac{e^2}{\kappa|\vec{r}|} \nonumber\\
&+&\frac{e}{c}\left[\frac{1}{2}\vec{B}\times\vec{R}\cdot\dot{\vec{r}}
+\frac{1}{2}\vec{B}\times\vec{r}\cdot\dot{\vec{R}}+
\frac{(m_1-m_2)}{M}\frac{1}{2}\vec{B}\times\vec{r}\cdot\dot{\vec{r}}\right]\label{L2}
\end{eqnarray}
where $M=m_1+ m_2$ is the mass of the center of mass and
$\mu=m_1m_2/M$ is the reduced mass. From now on, we consider
$m_1=m_2$. In contrast to the case of two identical
particles\cite{ajp65,pla269}, the present Lagrangian function is
similar to particles with opposite charges because it is a
nonseparable problem; the motion of the center of mass is coupled to
the relative variable\cite{ajp65}. The conjugate momentum for the
center of mass is given by
\begin{equation}
\vec{P}_{\vec{R}}=\frac{\partial
L(\vec{R},\vec{r};\dot{\vec{R}},\dot{\vec{r}})}{\partial
\dot{\vec{R}} }= M\dot{\vec{R}}+\frac{e}{2c}\vec{B}\times\vec{r},
\label{PR}
\end{equation}
which depends on  center of mass and relative motions. From deriving
the equations of motion we have
\begin{equation}
\dot{\vec{P}}_{\vec{R}}=\frac{\partial
L(\vec{R},\vec{r};\dot{\vec{R}},\dot{\vec{r}})}{\partial\vec{R}}=-\frac{e}{2c}\vec{B}\times\dot{\vec{r}}
\label{P1}
\end{equation}
If we integrate the Eq.(\ref{P1}), and it compares with the
Eq.(\ref{PR}), we obtain the first constant of motion
\begin{equation}
M\dot{\vec{R}}+\frac{e}{c}\vec{B}\times\vec{r}\equiv\vec{C},
\label{P2}
\end{equation}
and the first equation of motion is given by
\begin{equation}M\ddot{\vec{R}}=-\frac{e}{c}\vec{B}\times\dot{\vec{r}}.\end{equation} We obtain for the relative
vector conjugate momentum
\begin{equation}
\vec{p}_{\vec{r}}=\frac{\partial
L(\vec{R},\vec{r};\dot{\vec{R}},\dot{\vec{r}})}{\partial
\dot{\vec{r}} } = \mu\dot{\vec{r}}+\frac{e}{2c}\vec{B}\times\vec{R},
\label{Prr}
\end{equation}
thus, the force is given by
\begin{equation}
\dot{\vec{p}}_{\vec{r}}=\frac{\partial
L(\vec{R},\vec{r};\dot{\vec{R}},\vec{r})}{\partial {\vec{r}} } = -
\frac{e}{c}{\vec{B}}\times\dot{\vec{R}}-\frac{e^2}{\kappa{|\vec{r}|}^2}\hat{e_r}.
\label{Ppr}\end{equation} If we derive the Eq.(\ref{Prr}) in time,
and we compare with Eq.(\ref{Ppr}), we can obtain the second
equation of motion
\begin{equation}
\mu\ddot{\vec{r}}=-\frac{e}{c}\vec{B}\times\dot{\vec{R}}-\frac{e^2}{\kappa{|\vec{r}|}^2}\hat{e_r}.
\label{Em2}
\end{equation}
The second constant of motion is $E$, the energy of the system. The
general way to calculate the energy from a knowledge $L$ is
\begin{equation}
E=E(\vec{R},\vec{r};\dot{\vec{R}},\vec{r}) =
\vec{P}_{\vec{R}}\cdot\dot{\vec{R}}+\vec{p}_{\vec{r}}\cdot\dot{\vec{r}}
-L(\vec{R},\vec{r};\dot{\vec{R}},\vec{r}),\label{H1}
\end{equation}
and if we take Eq.(\ref{L2}) and we replace it into Eq.(\ref{H1}) we
can explicitly obtain

\begin{equation}
E=\frac{1}{2}M{\dot{\vec{R}}}^2+\frac{1}{2}\mu{\dot{\vec{r}}}^2-\frac{e^2}{\kappa|\vec{r}|},
\label{Ener}
\end{equation}
Putting Eq.(\ref{P2}) in Eq.(\ref{Ener}), we obtain an expression
for the energy in terms of the relative variable only.
\begin{equation}
E=\frac{1}{2}\mu{\dot{\vec{r}}}^2-
\frac{1}{2M}\left|\frac{e}{c}\vec{B}\times\vec{r}-\vec{C}\right|^2-\frac{e^2}{\kappa|\vec{r}|}.
\label{Ener2}
\end{equation}
A particular case has been discussed in Ref.\cite{Pursey}, where the
motion of the center of mass of the dipole is restricted to the $Z$
direction and rotational degrees of freedom of the dipole to the
$XY$ plane and the magnetic field is confined to the $Y$ direction.
Hence, we recover the classical behavior of the system from basic
equations of the present formulation.

\section{Motion in a Perpendicular Plane}
We will choose the most natural motion of electric charges in a
magnetic field. Let us consider a confined motion to two dimensions
in a perpendicular plane to the magnetic field. In this way, we
restrict the degrees of freedom of the center of mass and the
relative motion to the plane XY, the direction of the uniform
magnetic field is chosen to be in the $Z$ direction, this is,
$\vec{B}=B\hat{z}$. The separation between charges is constant, and
$|\vec{r}|= a$ is the linear size of the dipole. Let us define
variables and parameters in terms of the dimensionless units:
$\vec{\rho}=\vec{r}/a$ , $\vec{\xi}=\vec{R}/a$,
$\dot{\vec{\rho}}=\dot{\vec{r}}/\omega_c a$,
$\dot{\vec{\xi}}=\dot{\vec{R}}/\omega_c a$, if $E_0=\frac{1}{2}M
{a}^2{\omega_c}^2$, then $\varepsilon = E/E_0$,
$\dot{\varphi_c}=\dot{\varphi}/\omega_c$,
$\dot{\theta_c}=\dot{\theta}/\omega_c$ , where $\omega_c= eB / Mc$
is the cyclotron frequency and $a$  is the size of a dipole. We
define $\varepsilon_c\equiv e^2 / E_0\kappa a$ as the ratio between
the Coulomb energy and magnetic energy. In summary, distances have
been defined in terms of $a$ and the time in term of $1/w_c$.

From Eq.(\ref{P2}) and considering the dimensionless parameters, we
define a dimensionless constant of motion as follows
\begin{equation}
\dot{\vec{\xi}}+\hat{z}\times\vec{\rho}=\hat{z}\times\vec{\Upsilon}
\label{Upsi},
\end{equation}
where $\vec{\Upsilon}$ is a vector constant of motion and
$\vec{\rho}$ represents a position in the perpendicular plane $XY$
to the magnetic field. In addition, from Eq.(\ref{Ener}) and
considering the dimensionless parameters, we write the energy as a
dimensionless equation
\begin{equation}
\varepsilon=|\dot{\vec{\xi}}|^2 +
\frac{1}{4}|\dot{\vec{\rho}}|^2-\varepsilon_c \label{epsi0}.
\end{equation}
Now, if we replace Eq.(\ref{Upsi}) into Eq.(\ref{epsi0}), we write
the energy as a function of a single variable; this is,
\begin{equation}
\varepsilon=\frac{1}{4}|\dot{\vec{\rho}}|^2 +
|\vec{\rho}-\vec{\Upsilon}|^2-\varepsilon_c \label{epsi}.
\end{equation}
The dimensionless effective potential is given by
\begin{equation}
V_{eff}=|\vec{\rho}-\vec{\Upsilon}|^2-\varepsilon_c\label{Veff}.
\end{equation}
By following, if the linear size of the dipole is the unit of the
distance (i.e., $|\vec{\rho}|=1$), we can write
\begin{equation}
V_{eff}=1+ {\vec{\Upsilon}}^2 - 2\Upsilon
\cos\Psi-\varepsilon_c,\label{Veff2}
\end{equation}
where $\Psi$ represents the angle between the line that joins the
charges of the dipole and the vector constant $\vec{\Upsilon}$.
Thus, the effective potential $V_{eff}$ is a periodic function of
$\Psi$. Finally, we write an explicit expression for the
dimensionless energy given by
\begin{equation}
\varepsilon= \frac{1}{4}\dot{\Psi}^2+1+ {\vec{\Upsilon}}^2 -
2\Upsilon \cos\Psi-\varepsilon_c.\label{epsi2}
\end{equation}
\begin{figure} \centering
\includegraphics[angle=90,width=0.7\textwidth]{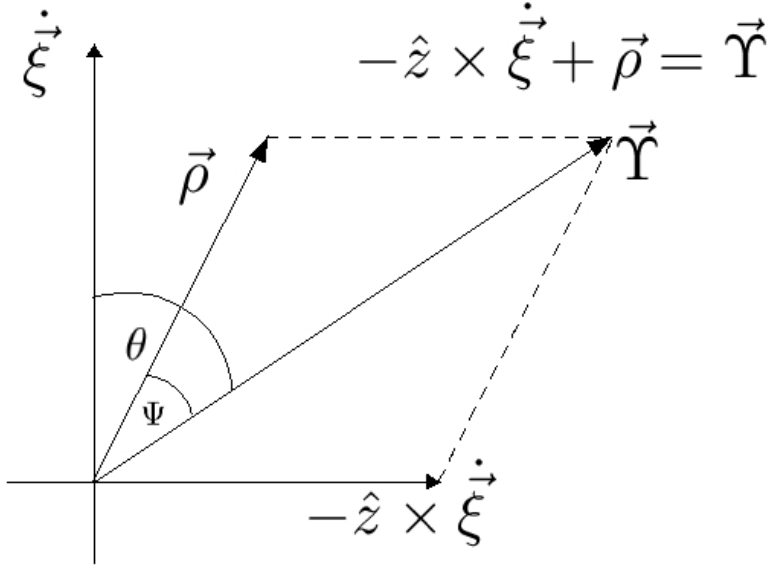}
\caption{\label{fig1} An illustrative picture is shown for the angle
$\Psi$. In addition, other parameters related to the geometry and
the orientation of the dipole are depicted.}
\end{figure}
In addition, in the Figure (\ref{fig1}), angle $\Psi$, dimensionless
relative variable and the velocity of the center of mass are
depicted at fixed time. Several parameters related to the geometry
and the orientation of the dipole are shown.

\subsection{Bound states}
According to previous considerations, we know that the effective
potential is a periodic function of the angle $\Psi$. If the energy
$\varepsilon$ of the system lives between two extreme values,
$V_{eff}^{min}=\left((1-\Upsilon)^2-\varepsilon_c\right)$ and
$V_{eff}^{max}=\left((1+\Upsilon)^2-\varepsilon_c\right)$, the
motion is bounded by two turning points. The value of total energy
in such point $\varepsilon = V_{eff}(\rho)$ gives the limit of the
oscillation.

Hence, from the Eq.(\ref{epsi2}) and by considering that
$\dot{\vec{\rho}}\cdot\hat{\Psi}=d{\Psi}/d\tau $, we obtain by
direct integration
\begin{equation}
\Delta\tau(\Psi)=\tau(\Psi)-\tau(\Psi_0)=\frac{1}{2}\int_{\Psi_0}^{\Psi}\frac{d\Psi'}{[\varepsilon+\varepsilon_c-1-{\Upsilon}^2+2\Upsilon
\cos\Psi']^{1/2}}, \label{DT}
\end{equation}
where $\tau=\omega_ct$, which we write in terms of an incomplete
elliptic integral of the first kind as follows
\begin{equation}
\Delta\tau(\Psi)=\frac{1}{2}F\left[\arcsin\left(\frac{\sin\Psi/2}{\sin\Psi_\varepsilon/2}\right)\setminus
\frac{\Psi_\varepsilon}{2}\right]
\end{equation}
where $\tau=\omega_ct$

The typical oscillation of the variable $\Psi$ for bound states
represents a bound rotation between two limiting angles,
$\pm\Psi_\varepsilon$, the classical turning points. That limiting
angles are given by
\begin{equation}
\Psi_\varepsilon =
\arccos(\frac{\varepsilon+\varepsilon_c-{\Upsilon}^2-1}{2\Upsilon}),\label{Psi}
\end{equation}
From the Eq.(\ref{epsi}), the energy of the system determines the
amplitude of the the oscillation. According to the Eq.(\ref{Prr}),
if the dipole enters to the magnetic field without an angular
relative velocity, then the energy only depends on the initial
orientation of the dipole. From the Eq.(\ref{DT}) and the
Eq.(\ref{Psi}) the period of the motion can be given by
\begin{equation}
\tau_P=\frac{1}{\sqrt{2\Upsilon}}\int_{-\Psi_\varepsilon}^{\Psi_\varepsilon}\frac{d\Psi'}{[\cos\Psi'-\cos\Psi_\varepsilon]^{1/2}}.
\end{equation}
Now, we use the definition of the complete elliptic integral of the
first kind to write the period of the motion
\begin{equation}
\tau_P= \frac{2}{\sqrt{\Upsilon}}
K\left[\sin\frac{\Psi_\varepsilon}{2}\right].
\end{equation}
In addition, we can use the definition of the hypergeometric
function\cite{Arfken} to present the same, as follows,
\begin{equation}
\tau_P= \pi \frac{1}{\sqrt{\Upsilon}}\;\;\; {}_{2}{\cal{F}}_{1}
\left[\frac{1}{2},\frac{1}{2},\sin\frac{\Psi_\varepsilon}{2}\right].
\end{equation}
Using the approximation of small oscillations, i.e., for energies
near to the point of the stable equilibrium $V_{ef}^{min}$, the
dimensionless frequency $\omega_P$ is given by
\begin{equation}
\omega_P=\frac{\sqrt{\Upsilon}}{2}
\end{equation}
Clearly, this frequency depends on $\Upsilon$, which corresponds to
the initial orientation and the initial velocity of the center of
mass from the Eq.(\ref{Ppr}).

\begin{figure} \centering
\includegraphics[angle=270,width=0.8\textwidth]{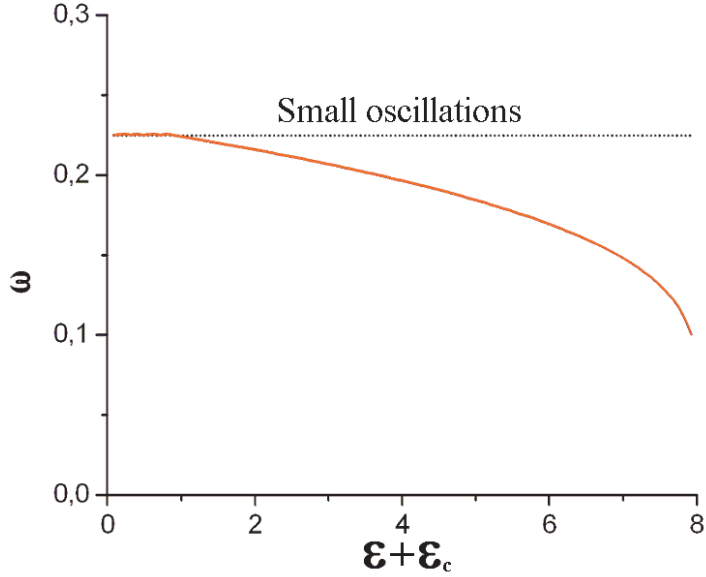}
\caption{\label{fig2} The frequency is depicted as a function of
energy $\varepsilon+\varepsilon_{c}$. The exact value for the
frequency coincides with the value in the aproximation of small
oscillations in the limit $\varepsilon\rightarrow-\varepsilon_c$}
\end{figure}

In the Figure~\ref{fig2} is depicted the relation between the
dimensionless frequency and the dimensionless energy. In the
approximation of small oscillations, we can get the effective
potential as a quadratic function of the angle $\Psi$. This is, the
period of such oscillations is independent of its amplitude. The
exact value for the frequency $\omega$ for
$\varepsilon+\varepsilon_c \rightarrow 0$ approaches to the
analytical value for the frequency $\omega_P$ from small
oscillations.

\subsection{Trapped States}
The classical trapped states are defined when the mean value of the
center of mass velocity is zero. From Eq.(\ref{Upsi}) we obtain
\begin{equation}
\langle \dot{\vec{\xi}}\rangle = \frac{1}{\tau}\int_0^{\tau}
d\tau'(-\hat{z}\times\vec{\rho}+\hat{z}\times\vec{\Upsilon})=0
\label{trap}
\end{equation}
where $\tau$ is the dimensionless period. The present condition
represents a closed motion. We expect to characterize a family of
unbounded states, without classical turning points, which can
satisfy the previous restriction. By following, we write by
compounds,

\begin{eqnarray}
\langle \dot{\vec{\xi_{\parallel}}}\rangle &=&
-\frac{1}{\tau}\int_0^{\tau}  d\tau'\cos\theta=0 \label{int1}\\
\langle \dot{\vec{\xi_{\perp}}}\rangle &=&
\frac{1}{\tau}\int_0^{\tau} d\tau'(-\sin\theta+\Upsilon
)=0,\label{int2}
\end{eqnarray}
where $\theta$ is the angle between $\dot{\vec{\xi}}$ and
$\vec{\Upsilon}$ (see Figure~\ref{fig1}) and the symbols
$\dot{\vec{\xi_{\parallel}}}$ and $\dot{\vec{\xi_{\perp}}}$ are
referred to the parallel and perpendicular parts of
$\dot{\vec{\xi}}$ respecting to the vector $\vec{\Upsilon}$.
Furthermore, we take into account that $\vec{\Upsilon}= \Upsilon
\hat{\Upsilon}$, where $\hat{\Upsilon}$ is a unit vector. We see
that if $\Upsilon=0$, Eq.(\ref{int1}) and Eq.(\ref{int2}) are simply
satisfied, the classical motion is trapped for any dimensionless
energy $\varepsilon > 1-\varepsilon_c$, , we obtain from
Eq.(\ref{epsi2}) the frequency of the trapped state that is given by
$\dot{\Psi}=2 \sqrt{\varepsilon+\varepsilon_c-1}$. In addition, we
can emphasize that if $\Upsilon \neq 0$, Eq.(\ref{int1}) is always
satisfied; however, the Eq.(\ref{int2}) imposes that $-1\leq
\Upsilon\leq 1$, while if $|{\Upsilon}|> 1$ the trapping should not
be possible.

\section{Summary} We have proposed to introduce a description of the classical dynamics of
rotors such as electric dipoles in an external magnetic field. Our
treatment does not consider radiation effects; however, the solution
is nontrivial; therefore, we believe the present discussion would be
pedagogically useful as a special topic of a standard course of
classical mechanics.

In the present application, we have considered the motion in a
perpendicular plane to a uniform magnetic field. The motion axis
coincides with the direction of the magnetic field. We find the
equations and constants of motion through the Lagrangian formulation
in the symmetric gauge. Our results are conveniently presented in
dimensionless variables. In the present view, bound states are
obtained for a periodic potential with initial conditions, which
satisfy Eq.(\ref{Upsi}) and Eq.(\ref{epsi0}). We are able to write a
simple expression for the frequency of motion,
$\omega_P=\sqrt{\Upsilon}/{2}$, in the approximation of small
oscillations. Trapped motion is obtained as a particular case of the
motion of the center of mass with an additional restriction, which
satisfy Eq.(\ref{trap}), where there are no turning points. The
definition of trapped states leads to a special range of values of
the constant of motion, this is $|{\Upsilon}|< 1$.

\section*{Acknowledgments}
This work has benefitted from partial support from FONDECYT 1051075.
Authors are very indebted to H. Alarc\'on for its helpful comment on
the draft version of this paper.
\newline


\begin{thebibliography}{88}
\bibitem{science306} Hern\'andez J V, Kay E R and Leigh D A 2004 A Reversible
Synthetic Rotary Molecular Motor {\em Science} {\bf 306} 1532-1536

\bibitem{science281} Gimzewski J K, Joachim C, Schlittler R R,
Langlais V, Tang H and Johansen I 1998 Rotation of a Single Molecule
Within a Supramolecular Bearing {\em Science} {\bf 281} 531-533

\bibitem{nat406_608} Bermudez V, Capron N, Gase T, Gatti F G, Kajzar F, Leigh D A, Zerbetto F and Zhang S
2000 Influencing intramolecular motion whith an alternating electric
field {\em Nature} {\bf 406} 608-611

\bibitem{nat406_605} Yurke B, Turberfield A J, Mills A P, Simmel F C and Neumann J L 2000 A DNA-fuelled molecular
machine made of DNA {\em Nature} {\bf 406} 605-608

\bibitem{nat401_150} Kelly T R, De Silva H and Silva R A 1999 Unidirectional rotary motion in a
molecular system {\em Nature} {\bf 401} 150-152

\bibitem{nat401_152} Koumura N, Zijlstra R W J,  van Delben R A,
Harada N and Feringa B L 1999 Light-driven monodirectional molecular
rotor {\em Nature} {\bf 401} 152-154

\bibitem{BJP29} Guimar\~aes P and Oliveira I S 1999 Classical and Quantum Mechanics of a Charged Particle in Oscillating
Electric and Magnetic Fields, {\em Braz. J. of Phys.} {\bf 29}
541-546

\bibitem{pra59} Spavieri G 1999 Quantum effect for an electric dipole {\em Phys. Rev.}  {\bf A 59} 3194-3199

\bibitem{ajp65} Curilef S and Claro F 1997 Dynamics of two interacting particles in
a magnetic field in two dimensions {\em Am. J. Phys.} {\bf 65}
244-250

\bibitem{pla269} Truong T and Bazzali D 2000 Exact low-lying states of two interacting equally charged particles in a magnetic field {\em Phys. Lett.} {\bf A 269} 186-193
\bibitem{jpa2} Taut M 1999 Two particles whit opposite charge in a homogeneous magnetic field:
particular analytical solutions of the two-dimensional Schrodinger
equation {\em J. Phys A: Math. Gen.} {\bf 32} 5509-5515
\bibitem{Pursey} Pursey D L, Sveshnikov N A, Shirokov A M 1998 Electric dipole in a magnetic field: Bound states without
classical turning points {\em Theor. and Math. Phys.} {\bf 117},
1262-1273
\bibitem{Arfken} Arfken G B and Weber H J 2001 Mathematical Methods for
Physicists, Fifth Edition, Academic Press
\end{thebibliography}
\end{document}